\documentclass{sig-alternate-05-2015}
\usepackage[utf8x]{inputenc}
\usepackage[T1]{fontenc}
\usepackage{amsmath}
\usepackage{amssymb}
\usepackage{subfig}
\usepackage{caption}\DeclareCaptionType{copyrightbox}
\usepackage{balance}
\usepackage{hyperref}
\usepackage{booktabs}

\usepackage[usenames,dvipsnames]{color}
\usepackage{epstopdf}
\usepackage{longtable}
\usepackage{tabularx}
\usepackage{afterpage}
\usepackage{booktabs}
\usepackage{xcolor}
\usepackage{enumitem}
\usepackage{amsfonts}
\usepackage{scalerel}
\usepackage{tikz}
\usepackage{url}
\usepackage{multirow}

\usepackage[pass]{geometry}

\usepackage{flushend} 

\title{Simplified Relative Citation Ratio \\ for Static Paper Ranking}

\subtitle{UFMG/LATIN at WSDM Cup 2016 \vspace{5pt}}

\numberofauthors{6} 
\author{
\alignauthor Sabir Ribas\\
       \affaddr{CS Dept, UFMG}\\
       \affaddr{Belo Horizonte, Brazil}\\
       \email{sabir@dcc.ufmg.br}
\alignauthor Alberto Ueda\\
       \affaddr{CS Dept, UFMG}\\
       \affaddr{Belo Horizonte, Brazil}\\
       \email{ueda@dcc.ufmg.br}
\alignauthor Rodrygo L. T. Santos\\
       \affaddr{CS Dept, UFMG}\\
       \affaddr{Belo Horizonte, Brazil}\\
       \email{rodrygo@dcc.ufmg.br}
\and  
\alignauthor Berthier Ribeiro-Neto\\
       \affaddr{CS Dept, UFMG \& Google Inc}\\
       \affaddr{Belo Horizonte, Brazil}\\
       \email{berthier@dcc.ufmg.br}
\alignauthor Nivio Ziviani\\
       \affaddr{CS Dept, UFMG \& Zunnit Tech}\\
       \affaddr{Belo Horizonte, Brazil}\\
       \email{nivio@dcc.ufmg.br}
}

\date{\today}

\newfont{\mycrnotice}{ptmr8t at 7pt}
\newfont{\myconfname}{ptmri8t at 7pt}
\let\crnotice\mycrnotice%
\let\confname\myconfname%

\CopyrightYear{2016} \setcopyright{acmcopyright}
\conferenceinfo{WSDM'16,}{February 22--25, 2016, San Francisco, CA, USA.}
\isbn{978-1-4503-3716-8/16/02}\acmPrice{\$15.00}
\doi{http://dx.doi.org/XXXX.XXXX} 

\begin{document}

\newcommand{\citep}[1]{\cite{#1}}

\maketitle

\begin{abstract}
Static rankings of papers play a key role in the academic search setting.
Many features are commonly used in the literature to produce such rankings, some examples are citation-based metrics, distinct applications of PageRank, among others. 
More recently, learning to rank techniques have been successfully applied to combine sets of features producing effective results.
In this work, we propose the metric S-RCR, which is a simplified version of a metric called Relative Citation Ratio --- both based on the idea of a co-citation network. 
When compared to the classical version, our simplification S-RCR leads to improved efficiency with a reasonable effectiveness.
We use S-RCR to rank over 120 million papers in the Microsoft Academic Graph dataset. 
By using this single feature, which has no parameters and does not need to be tuned, our team was able to reach the 3rd position in the first phase of the WSDM Cup 2016.
\end{abstract}



\section{Ranking Papers}\label{sec:introduction}


Finding the most relevant papers of a field of knowledge is a task with many motivations. From the researcher's perspective, it is important for instance to quickly discern the papers with major impact in his/her study area from those with less relevance. On the other hand, from an academic search engine perspective, a common task is to present the papers by using rankings, which demands a sort criterion as relevance. Also, establishing a relative order of importance of papers could help in other tasks such as providing grants or research awards for individual researchers and graduate programs.
The problem of ranking papers was addressed in the WSDM Cup 2016, a competition that brought together 32 research teams from all over the world.

\subsection{The Competition --- WSDM Cup 2016}

The WSDM Cup Ranker Challenge\footnote{\url{http://wsdmcupchallenge.azurewebsites.net}} was created by the WSDM\footnote{\url{http://www.wsdm-conference.org/2016/}} organizers and supported by Microsoft Research.

\subsubsection{The Task}

The task for each competitor in WSDM Cup 2016 was to provide the best static rank values for publication entities in the Microsoft Academic Graph\footnote{\url{http://research.microsoft.com/en-us/projects/mag/}}~(MAG)~\cite{mag-www}. The goal behind it was to assess the query-independent importance of academic papers.

\subsubsection{Evaluation}

The evaluation in WSDM Cup 2016 was conducted in two phases, as we now describe.
%
%
During Phase 1, submissions were scored based on the agreements with human judgement data. A group of Computer Science researchers were invited by the organizers to conduct a pairwise ranking of papers in the fields they actively conduct research. The pairwise judgement data were then randomly segregated into an Evaluation and a Test set. Submissions during Phase 1 were automatically scored against the Evaluation set and added to a public leaderboard 
that was sorted based on the percentage of agreements with the judgement data.
At the end of Phase 1, the most recent submission from each team was evaluated against the Test set and the scores (ranked by the percentage agreements with the Test set) were announced to the leaderboard. 


The top eight teams on the leaderboard at the end of Phase 1 were invited to participate in Phase 2. Each participant of Phase 2 was asked to re-run the algorithms over an updated graph and to submit the final rank values. Phase~2 of the Challenge was conducted by Microsoft Research in cooperation with Bing. Each of the finalist results was applied to Bing search results and powered the ranker used by Bing for academic queries. 
%


\subsection{This Report}

In this paper, we report the participation of our team, named UFMG/LATIN, in the WSDM Cup 2016.
Before getting into the final model, we performed a set of tests considering distinct approaches, which include distinct citation-based metrics, PageRank, among others.
Our final approach was based on a simplified version of a metric called Relative Citation Ratio~\citep{hutchins2015relative}. In here, we describe this metric, how and why we choose to run a simplified version.

The remainder of this paper is organized as follows. In Section~\ref{sec:literature}, we present the related works on paper rankings. In Section~\ref{sec:rcr}, we describe the RCR metric as well as our simplified version, the S-RCR. In Section~\ref{sec:features}, we discuss some additional techniques we applied to rank papers. In Section~\ref{sec:experiments}, we report our experiments. Finally, in Section~\ref{sec:discussion}, we provide a final discussion and concluding remarks.




\section{Literature on Paper Rankings}\label{sec:literature}

The most common approach to produce paper rankings in the literature is by using citation-based metrics. These metrics provide a natural way to reason about the relative quality of academic entities, such as scientific papers, individual researchers and publication venues. 
One of the earliest metrics proposed to quantify academic impact was the Impact Factor \citep{garfield1955}. Since them, many alternatives have been proposed, including other citation-based metrics like the H-Index~\citep{hirsch2005}, random walks and machine learning techniques. 
%

Another common approach to rank academic entities is by considering the structural information.
The structure of the citation network can be used to produce academic rankings by applying random walk techniques, such as the PageRank~\citep{page98pagerank}. 
A natural approach is to apply random walks in the paper-paper citation network. However, some authors also apply random walks in heterogeneous graphs.
In~\citep{ribas2015random}, for example, the authors propose a novel random walk model to identify the most reputable entities of a domain based on a conceptual framework of reputation flows. 

Another concept that is worth mentioning is the Altmetrics movement, which points out the need for novel evaluation metrics as alternative to classic citation-based metrics. According to Piwowar et al.~\citep{piwowar2013altmetrics}, citation-based metrics are useful, but not sufficient to evaluate research. 
In particular, they observe that citations are slow --- their main argument is the fact that a paper's first citation can take years. 

Learning to rank techniques~\citep{l2r} have been used over the last few years to improve the quality of rankings by effectively combining multiple sources of evidence. The large amount of available features related to some ranking tasks motivates the adoption of learning to rank methods in distinct contexts, including in academic search. 

Our approach is inspired by the work of Hutchins et al.~\citep{hutchins2015relative}, which proposes a metric called Relative Citation Ratio. It is a paper-level and field-independent score that provides an alternative to classic citation-based metrics to identify influential papers. 
They show that the rankings produced by their metric strongly correlate with the opinions of experts in biomedical research and suggest that the same approach should be generally applicable in all areas of science. 

While complex models, such as heterogeneous random walks or learning to rank methods, are able to produce effective results, in this work, we investigate a single well-designed feature to rank papers given its citation network. 

\section{Relative Citation Ratio}\label{sec:rcr}

In this section, we describe the metric we propose, the S-RCR, which is a simplification of a metric called Relative Citation Ratio (RCR)~\citep{hutchins2015relative}. 
Before describing S-RCR it is worth reasoning about the basic concepts of the original RCR metric.
The RCR metric is based upon the idea of using the co-citation network of each paper to normalize it in terms of time and area of study, by calculating an expected citation rate of the target paper from the aggregated citation behavior of its neighborhood. Basically, this strategy consists of computing the average citation rate of this neighborhood which is used as the RCR denominator; the numerator is the citation rate of the target paper. 

\subsection{Co-citation Network}\label{sec:neighborhood}

\begin{figure}[h]
   \centerline{\includegraphics[scale=0.55]{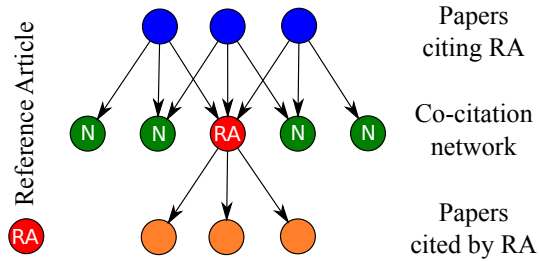}}
   \caption{Schematic view of a co-citation network~\cite{hutchins2015relative}}
   \label{fig:rcr}
\end{figure}

The basis of the RCR metric and our proposed simplification is the notion of co-citation network. Hutchins et al.~\cite{hutchins2015relative} define this co-citation network as the \textit{papers' area of influence}.
As described in \cite{hutchins2015relative}, when a paper is first cited, the other papers appearing in the reference list along with this paper comprise its co-citation network, see Figure~\ref{fig:rcr}.
As the paper continues to be cited, the papers appearing in the new reference lists alongside it are added to its co-citation network.
This network provides a dynamic view of the paper's field of research, taking advantage of information provided by the experts who have found the study useful enough to cite.
The co-citation network of a paper can be viewed as a representative sample of its area of research allowing us to perform a reasonable cross-field evaluation of papers.


In Figure~\ref{fig:rcr}, we present the schematic view of a co-citation network for the RCR computation. The Reference Article (RA, in red) cites previous papers from the literature (in orange) and other papers (in blue) cite the RA. The co-citation network (or neighborhood) of the Reference Article is the set of papers (in green) that appear alongside the RA.

\subsection{Article Citation Ratio}\label{sec:acr}

The Article Citation Ratio (ACR) of a given paper $p$ is defined as:

\begin{equation}
\text{ACR}(p) = \frac{\text{Citations}(p)}{\text{Age}(p)+1},
\end{equation}
where Citations$(p)$ is the total number of citations paper $p$ received since its publication and Age$(p)$ is the time in years since the publication date of paper $p$.

By keeping a counter of citations and paper ages, we can store the data in a hash table and compute the ACR of a paper $p$ in $\Theta(1)$ time complexity.

\subsection{The Simplified RCR}

To produce field independent rankings, RCR metrics normalize the ACR of a paper $p$ based on the information of its co-citation network. 
In the original version of RCR~\citep{hutchins2015relative}, this information is used through a complex normalization process. Computing the original RCR of a single paper $p$ depends on performing a linear regression on its co-citation network using the journal citation ratio~\citep{garfield1972} of the venues $p$'s neighbors were published. 
For a full description of the RCR metric, we refer the reader to the work of Hutchins et al.~\citep{hutchins2015relative}.

In particular, we define the Simplified Relative Citation Ratio (S-RCR) of a given paper $p$ as follows:

\begin{equation}\label{eq-srcr}
\text{S-RCR}(p) = \frac{\text{ACR}(p)}{(1/|N_p|) \sum_{p' \in N_p}{ACR(p')}},
\end{equation}
where ACR$(p)$ is the Article Citation Ratio (see Section~\ref{sec:acr}) of paper $p$ and $N_p$ is the set of neighbors of paper $p$ (see Section~\ref{sec:neighborhood}). 
Similarly to the classic RCR, the numerator is the ACR of paper $p$ and 
the denominator acts as a normalizer, forcing the ACR of paper $p$ to be relative to its neighbors. 

The main difference between our proposal and the original RCR is its normalization step, which, in our metric, is much simpler. 
Specifically, we normalize the ACR of a paper $p$ by the average of the ACR values of $p$'s neighbors.
While this simplification is briefly mentioned in the original paper, the authors discard it for its numerical limitations, e.g., for papers with no neighbors. In contrast, we overcome these limitations by smoothing Eq.~(\ref{eq-srcr}) via additive smoothing.

If, for example, a target paper has the same ACR than the average of its neighborhood, its S-RCR value is equal to~1. An S-RCR higher than 1 indicates that the paper has a relevance signal stronger than its co-citation network. Similarly, an S-RCR value lower than 1 indicates low relevance of the paper within its neighborhood. 

Since the ACR function can be computed in $\Theta(1)$, the time complexity to compute the S-RCR of a given paper $p$ is $\Theta(|N_p|)$, where $|N_p|$ is the neighborhood size of paper $p$.
This time complexity is lower than the time complexity of the classic RCR since, to compute the classic version of RCR, we need to perform a linear regression on $p$'s neighborhood. 
%

\section{Other Features}\label{sec:features}

Besides the S-RCR metric, we performed a set of tests using features that 
%
ended up not being used 
in our final ranking. These features include paper raw citation counts and normalizations, distinct aggregations of citation-based metrics of authors and publication venues, among others. 
We also tried some Random Walk techniques, like an application of PageRank on the paper citation graph and a Random Walk on a heterogeneous Pseudo-Tripartite graph composed by paper, author and venue nodes. In this last approach, there is an edge between an author $a$ and a paper $p$ if author $a$ is one of the authors of paper $p$. Also, there is an edge between paper $p$ and venue $v$ if paper $p$ was published in venue $v$. The interaction between papers was given by the paper citation graph. While we believe that a proper investigation of this last approach would lead to effective results, it depends on the calibration of the parameters needed to control the amount of probability mass between nodes of distinct types.
A plus of our final approach based on the proposed S-RCR metric is that it produces effective results without the need of any parameter tuning.
\section{Experiments}\label{sec:experiments}

In this section, we report some experiments performed by our team during the 1st phase of the WSDM Cup 2016. 

\subsection{Dataset Description}

The Microsoft Academic Graph (MAG)~\cite{mag-www} is a heterogeneous graph containing scientific publication records, citation relationships between publications -- as well as authors, institutions, journals, conferences, and fields of study. The file size (zipped) is 37GB and it contains individual information about more than 120 million papers. During the competition, three versions of this dataset were released. Here, we describe only the last one 
(version of Nov. 6, 2015).


\begin{table}[htbp]
\centering
\caption{Relevant statistics}
\begin{tabular}{|l|r|}
\hline
Papers with citation information    &  49,870,036 \\ \hline
Papers without citation information &  71,017,797 \\ \hline
Total number of papers              & 120,887,833 \\ \hline
Average neighborhood size           &         891 \\ \hline
\end{tabular}
\label{tab:dataset}
\end{table}


In Table~\ref{tab:dataset}, we present some statistics of the MAG dataset that are relevant to this work. Notice that a large fraction (59\%) of the papers in this dataset have no information on citations, that is, the paper can be represented as a node in the citation graph that has neither inlinks nor outlinks.
There are two possible reasons for this to happen. 
The first alternative is the zero degree (i.e., both in- and out-degree) is a true representation of reality, it is a paper that in fact does not receive any citation yet and does not cite any other paper.
The second alternative is due to the fact that any big repository offers an approximation of the reality, which also happens in the MAG dataset. Collecting and organizing a real-world dataset of such size is not a trivial task. In fact, it is a process that involves the treatment of huge amounts of semi-structured data, 
which usually causes some inconsistencies.

\begin{figure}[h]
   \centerline{
   \includegraphics[scale=0.5]{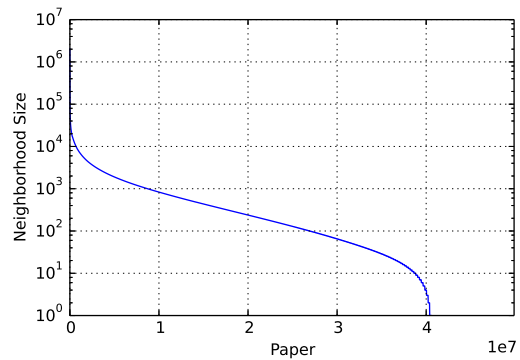}
   }
   \caption{Distribution of neighborhood sizes}
   \label{fig:neighbors}
\end{figure}

In Figure~\ref{fig:neighbors}, we plot the distribution of neighborhood sizes. 
To compute this distribution, we consider only the approximately 50 million papers that have information on citations. 
This figure helps us to characterize the neighborhood sizes of the graph. 
We notice that the neighborhood sizes follow a long tail distribution, there are many papers with just a few neighbors and few papers with large neighborhoods.

\subsection{Submissions}

Our team made many submissions in the first phase of the competition. In this section, we discuss some of them.
As mentioned previously, each submission is a ranking where each item is a paper ID and a probability of that paper being important. 
The only evaluation sign available was the score reported by the competition's page for each submission, a value between zero and one representing the quality of the submitted ranking --- the higher the better. While we know that the evaluation score is based on previously computed pair-wise expert judgments, the exact evaluation metric was not disclosed by the organizers. 

In Table~\ref{tab:leaderboard}, we present the evaluation scores our rankings received in the first phase of the competition and also the time elapsed to produce each submission file.

\begin{table}[htbp]
\centering
\caption{Our ranking scores in the 1st phase of WSDM Cup}
\begin{tabular}{|l|c|c|c|}
\hline
          & \multicolumn{2}{c|}{Leaderboard} &  \\ \cline{2-3}
Feature   & Public & Private & Time \\ \hline
Citations & 0.675 & -        & 0h08m \\ \hline
PageRank  & 0.687 & -        & 1h29m \\ \hline
ACR       & 0.685 & -        & 0h16m \\ \hline
S-RCR     & 0.697 & 0.671    & 1h50m \\ \hline
\end{tabular}
\label{tab:leaderboard}
\end{table}

A first guess for one who is somehow familiar with the problem is to count paper citations in order to rank them. It was our first submission and received the score of 0.675 in the public leaderboard.
Using PageRank is also a natural approach to rank papers in a citation graph. Our PageRank-based submission scored 0.687, which represents an improvement of 1.8\% over citation counts.
Before submitting the S-RCR, we submitted the ranking produced by its component ACR (see Section~\ref{sec:acr}). The ACR submission was scored higher than citation counts but lower than PageRank.
Finally, our highest scored submission was based on the S-RCR metric. Scoring 0.697 in the public leaderboard, it corresponds to an improvement of 3.3\% over citation counts. 

Since the ranking produced by S-RCR received the highest public score among our submissions, we used it as the final ranking of the first phase. As part of the competition, it was evaluated using a private test set and received the score of 0.671, which was sufficient to bring our team UFMG/LATIN to the 3rd position at the end of the first phase.
Eight teams were promoted to the second phase of the competition: the score of the 1st-placed team was 0.684, while the score of the 8th-placed team was 0.642.
It is noteworthy that our score dropped a little in the final evaluation of the 1st phase, from 0.697 to 0.671. 
Some teams experienced higher drops, apparently due to overfitting.

We ran our experiments on a machine with 64 GB RAM, 24 processors of 3.33GHz --- Intel(R) Xeon(R) CPU X5680 --- under Ubuntu 14.04.2 LTS. 
While we did not take advantage of the full computational power available (by not using parallelism in a 24-cored machine), the processing times were pretty low for a graph of such size. The time elapsed to produce our final submission file was of only 1h50m.

Critical parts of our approach, like graph processing, were implemented in C++, while other parts, like intermediate analysis or file treating/formatting, were implemented in Python --- Pandas and Jupyter were useful tools. 


\section{Discussion}\label{sec:discussion}

In this work, we have proposed the S-RCR metric and applied it to produce static rankings of academic papers in the Microsoft Academic Search dataset.
The interesting point here is that a single well-designed feature (which is a simplification of a more complex one) was able to produce effective results,
promoting our team to the 3rd place in the first phase of the WSDM Cup 2016 competition.
%
This fact reinforces the argument that feature engineering is at least as important as complex models, since we apply a single well-designed feature that leads to better results than complex models with the advantage of less tuning and less computational effort. Also, single features tend to be more interpretable.
%
%
A future direction that is worth investigating is the impact of using the S-RCR metric together with other features in learning to rank techniques.
Another direction is to study approaches to address ranking ties, specially how to break ties between papers with no information on citations.
Using reputation-based metrics \citep{ribas2015bigscholar,ribas2015random} seems to be a reasonable approach to address these issues.
%

\section*{ACKNOWLEDGEMENTS}
This work was partially sponsored by the Brazilian National Institute of Science and Technology for the Web (MCT/ CNPq 573871/2008-6) and the authors' individual grants and scholarships from CNPq and FAPEMIG. 


\balance
\bibliographystyle{abbrv}
\bibliography{ribas2016wsdmcup}

\end{document}